\documentclass[a4paper]{article}
\usepackage{INTERSPEECH2022}

\usepackage[colorinlistoftodos,prependcaption,textsize=small]{todonotes}
\usepackage{hyperref}
\usepackage{tipa}
\usepackage{gensymb}
\usepackage{adjustbox}
\usepackage{amsmath}
\usepackage{mathtools}
\usepackage{flushend}

\DeclareMathOperator*{\argmin}{argmin}
\usepackage{booktabs}
\usepackage{cite}

\title{Exploration strategies for articulatory synthesis of complex syllable onsets}
\name{Daniel R. van Niekerk$^1$, Anqi Xu$^1$, Branislav Gerazov$^2$, Paul K. Krug$^3$, Peter Birkholz$^3$, Yi Xu$^1$}
\address{
  $^1$Department of Speech, Hearing and Phonetic Sciences, University College London, UK\\
  $^2$Faculty of Electrical Engineering and Information Technologies, CMUS, Skopje, RN Macedonia\\
  $^3$Institute of Acoustics and Speech Communication, Technische Universit\"at Dresden, Germany
}
\email{d.vniekerk@ucl.ac.uk}

\begin{document}

\maketitle
\begin{abstract}
  High-quality articulatory speech synthesis has many potential applications in speech science and technology. However, developing appropriate mappings from linguistic specification to articulatory gestures is difficult and time consuming. In this paper we construct an optimisation-based framework as a first step towards learning these mappings without manual intervention. We demonstrate the production of syllables with complex onsets and discuss the quality of the articulatory gestures with reference to coarticulation.
\end{abstract}

\section{Introduction}
\label{sec:intro}

High-quality articulatory speech synthesis provides compelling possibilities for studying articulatory phonetics and constructing low-resource speech technologies. However, many of these scenarios require a mapping from linguistic specification, e.g. a sequence of phonetic symbols to articulatory gestures. To date, such mappings have been developed manually by developers or users of the particular synthesizer. \emph{VocalTractLab}~\cite{birkholz2005thesis, birkholz2013coart}, a state-of-the-art synthesizer that simulates a vocal tract based on magnetic resonance imaging (MRI) data, is a system capable of producing natural sounding speech~\cite{krug11intelligibility}. However, the increase in realism comes with an increase in time and expertise required to develop gestures and the increase in speech quality reflects subtle articulatory choices which makes it difficult to develop language or dialect-independent gestures manually.\footnote{Furthermore, gestural mappings will need to be revisited each time a new speaker model is introduced.} The result is that the set of available gestures will be, at best, incomplete or in the worst case only appropriate for a specific language.

This problem can be resolved by a procedure that learns the articulatory gestures to produce linguistically relevant utterances automatically. To be practical, the process should function without detailed articulatory phonetic information such as aligned MRI data or expert knowledge. This resembles the task of \emph{spoken language acquisition}, in general, and \emph{early vocal learning} in particular~\cite{jusczyk1997spokenlang}. In this paper we implement one of the central processes for autonomous learning of speech production, namely articulatory exploration or ``babbling'', as an auditory optimisation task~\cite{oller1988babbling}.

Using this simulation, we show that it is possible to discover the articulation of syllables with complex onsets by using a perceptual encoder trained on a general speech recognition corpus to provide auditory objectives. Furthermore we demonstrate the relative success of different exploration strategies and examine the nature of articulatory solutions in terms of coarticulation between consonant and vowel gestures.

This is a significant development towards an autonomous process for constructing linguistic-to-articulatory mappings for any language or dialect and provides a framework for investigating theoretical questions in articulatory phonetics and speech perception in vocal learning.


\section{Approach}
\label{sec:appr}

\begin{figure*}[h]
  \centering
  \includegraphics[width=0.8\linewidth]{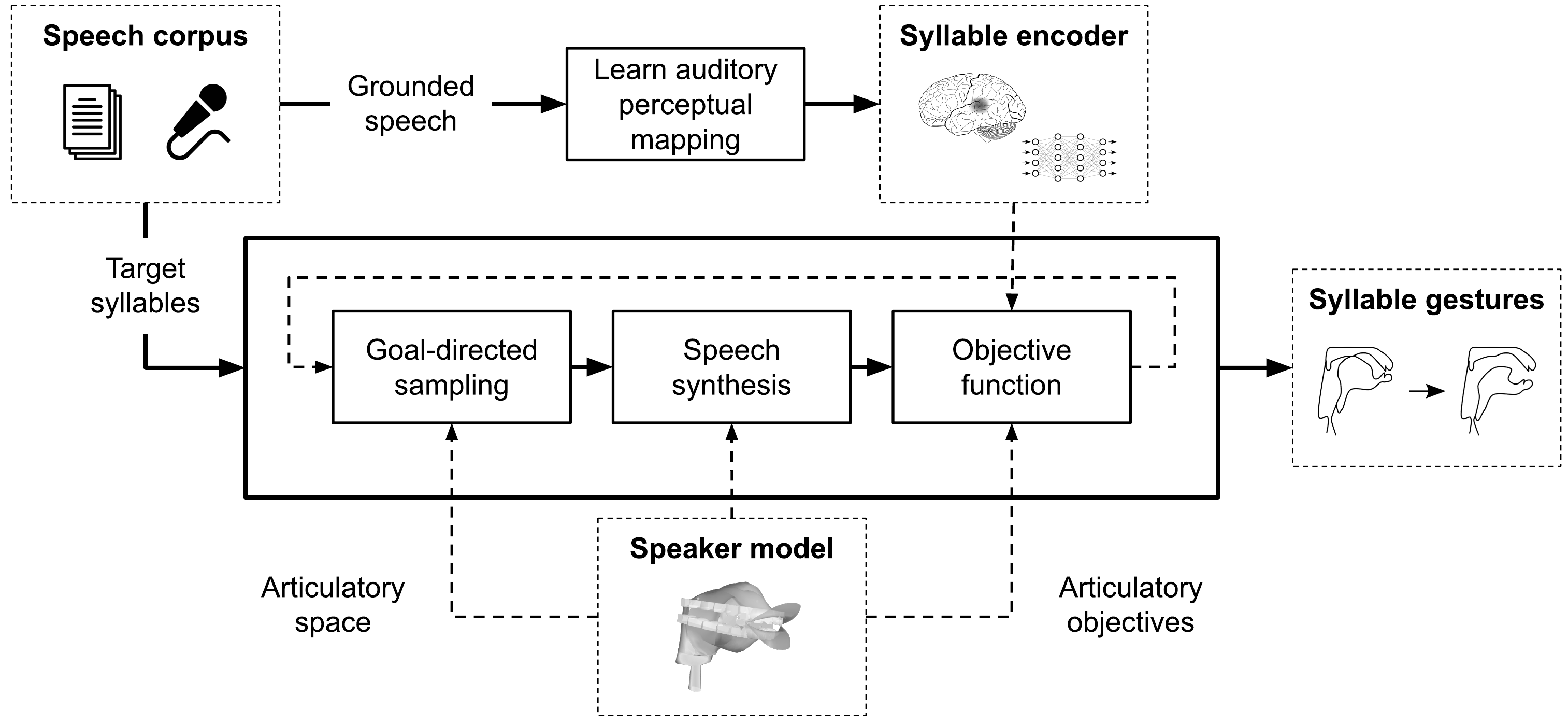}
  \caption{A process for discovery of linguistically relevant articulatory gestures. The central exploration task is goal-directed and relies on articulatory sampling, speech production, and auditory and articulatory objectives. Required input resources are a multi-speaker speech corpus and the vocal tract model.}
  \label{fig:optccv}
\end{figure*}

As in previous work~\cite{vniekerk2020cvopt} we formulate articulatory exploration as an optimisation task using \emph{VocalTractLab} (\emph{VTL}) to produce candidate utterances. However, in this work, novel mechanisms including well-motivated somatosensory specifications and a language-oriented auditory perceptual mapping form part of the objective function. The process of discovering linguistically relevant gestural mappings is illustrated in Figure~\ref{fig:optccv} and is briefly motivated in the following subsections.

\subsection{Articulatory exploration}
\label{sec:appr:opt}

Babbling during early vocal learning has often been simulated as a \emph{goal-directed or imitative} process, usually involving a set of auditory objectives \cite{bailly1997sensmot, rasanen2012phonla, rasilo2017learnvowel, pagliarini2021review, philippsen2021speechacq}. This type of exploration is also considered central to finding appropriate inverse models during sensorimotor learning in general~\cite{jordan1992distal, rolf2010goalbabble}.

We implement goal-directed articulatory exploration as the process of minimising auditory and somatosensory losses to discover a linguistically relevant utterance. The central block in Figure~\ref{fig:optccv} is the global optimisation task
\begin{equation}
  \label{eq:opt}
  \boldsymbol{u^*} = \argmin_{\boldsymbol{u} \in \boldsymbol{U}_\theta}{L(\boldsymbol{u}, \boldsymbol{q}, Q, \theta)}
\end{equation}
of finding the articulatory gestures $\boldsymbol{u^*}$ that minimise the loss function $L$, with $\boldsymbol{q}$ the combined auditory/articulatory goal, $Q$ the auditory perceptual mapping described in Section~\ref{sec:appr:rec} and $\theta$ the speaker vocal tract model. In this paper, we use the Tree-structured Parzen Estimator approach~\cite{bergstra2011hpopt} as the algorithm to drive the articulatory sampling.

\subsection{Auditory perceptual objectives}
\label{sec:appr:rec}

Most vocal learning simulations assume that auditory objectives are derived from the speech signal alone, i.e. acoustic imitation. This approach suffers from two issues: the \emph{speaker normalisation} and \emph{correspondence} problems. The speaker normalisation problem refers to the difficulty of finding linguistically correct utterances when comparing speech produced by different speakers; e.g. it is well known that formant frequencies vary systematically with speakers' vocal tract length and that this may affect speech recognition performance~\cite{waibel1997vtln}. The correspondence problem is one of associating articulatory gestures obtained for an acoustic reference to linguistic contexts~\cite{nehaniv2002corrprob, brass2005imit, philippsen2021speechacq}. We have argued elsewhere, based on the well-known finding that language-oriented speech perception precedes the onset of \emph{canonical babbling} in infants~\cite{kuhl2004evoc}, that these problems can be addressed by a language-oriented auditory perceptual mapping derived from linguistically grounded multi-speaker speech stimuli~\cite{vniekerk2022cvopt}.

\subsection{Articulatory objectives}
\label{sec:appr:art}

Articulatory objectives represent \emph{explicit objectives} that originate from non-auditory signals. In humans it is known that sighted individuals may benefit from visual information~\cite{murakami2015seeingu} and speakers may track the implementation of these objectives through somatosensory feedback~\cite{nasir2006somatosensory}. However, since we rely on a general optimisation algorithm with uniform priors instead of a physiologically motivated approach (as for example in~\cite{serkhane2007fcsim, nam2013apsim}) and the speaker model does not incorporate inertial and other relevant process measurements, articulatory objectives may also serve as an \emph{implicit mechanism} that regularises the solution space, i.e. resulting in more prototypical articulatory gestures.

In this paper we employ one set of \emph{somatosensory objectives} that could be derived from visual information: plosive consonants at the start of the syllable should form an oral closure and the vowel is associated with an open vocal tract. We also experiment with a \emph{regularisation objective} to induce intra-syllable coarticulation.

\subsection{Speech production}
\label{sec:appr:synth}

To produce articulatory trajectories, we use the target-approximation model (TAM)~\cite{xu2001tam} which has been adopted in \emph{VTL} to realise utterances represented by \emph{articulatory targets}~\cite{birkholz2007gestures}. The resulting parameterisation of the articulatory dynamics combined with simplifying assumptions of synchronisation~\cite{birkholz2011qtacv} has enabled the reliable discovery of simple CV syllables using derivative-free optimisation or even random sampling~\cite{xu2019icphs, vniekerk2020cvopt}. While previous works implemented coarticulation~\cite{xu2020sylsync, liu2021sylsync} by explicitly parameter tying~\cite{xu2019icphs, vniekerk2020cvopt}, this work tests the hypothesis by including coarticulation as an articulatory objective, which further reduces the explicit knowledge required in the process, as described in the next section.

\begin{table}[h]
  \centering
  \caption{Upper vocal tract parameters in \emph{VTL}.}
  \label{tab:articdims}
\begin{adjustbox}{width=0.9\columnwidth}
\begin{tabular}{@{}ll@{}}
\toprule
\multicolumn{1}{c}{\textbf{Parameter}} & \multicolumn{1}{c}{\textbf{Description}}                                   \\ \midrule
HX, HY                                 & Horiz. and vert. hyoid positions                                           \\
JX, JA                                 & Horiz. jaw position and jaw angle                                          \\
LP, LD                                 & Lip protrusion and vert. lip distance                                      \\
TTX, TTY                               & Horiz. and vert. tongue tip positions                                      \\
TBX, TBY                               & Horiz. and vert. tongue blade positions                                    \\
TCX, TCY                               & Horiz. and vert. tongue body centre positions                              \\
TRX, TRY                               & Horiz. and vert. tongue root positions                                     \\
VS, VO                                     & Velum shape and opening                                                               \\
TS1 – TS3                              & Tongue side elevation from the anterior to the \\
& posterior part of the tongue \\
\bottomrule
\end{tabular}
\end{adjustbox}
\end{table}

\section{Experimental setup}
\label{sec:exp}

\begin{table*}[t]
  \centering
  \caption{Identification rates (\%) for samples by vowel, onset, and complete syllable for each experimental condition ($5$ trials $\times 150$ syllable types $= 750$ samples per condition).}
  \label{tab:overall}
  \begin{adjustbox}{width=0.7\textwidth}
    \begin{tabular}{@{}l|rrrr|rrrr@{}}
      \toprule
      \multicolumn{1}{c}{} & \multicolumn{4}{c}{Without coarticulation objective} & \multicolumn{4}{c}{With coarticulation objective} \\
      \midrule
 & \multicolumn{1}{|l}{$C_1C_2V$} & \multicolumn{1}{l}{$V.C_1C_2$} & \multicolumn{1}{l}{$V.C_1.C_2$} & \multicolumn{1}{l}{$V.C_2.C_1$} & \multicolumn{1}{|l}{$C_1C_2V$} & \multicolumn{1}{l}{$V.C_1C_2$} & \multicolumn{1}{l}{$V.C_1.C_2$} & \multicolumn{1}{l}{$V.C_2.C_1$} \\
      syllable & 83.60 & 83.87 & 82.93 & \underline{69.47} & 79.97 & \textbf{85.07} & 81.15 & 
      \underline{70.53} \\
      vowel    & 91.33 & 93.07 & 92.27 & 90.53 & \underline{88.79} & 94.00 & 92.91 & 92.93 \\
      onset    & 91.07 & 90.53 & 90.27 & \underline{77.47} & 89.99 & 90.93 & 87.83 & \underline{76.80} \\
      \bottomrule
    \end{tabular}
  \end{adjustbox}
\end{table*}

\subsection{Implementation}
\label{sec:exp:impl}

The first step towards implementing the articulatory exploration process (Figure~\ref{fig:optccv}) to find CCV syllables is to construct the auditory perceptual mapping that produces syllable embeddings or percepts. For this purpose, we used the \emph{LibriSpeech} speech recognition corpus~\cite{panayotov2015librispeech} representing linguistically grounded speech. Vowel (V), consonant-vowel (CV) and CCV syllable onsets were extracted from the clean training set and used to train a recurrent neural network that encodes the audio to a low-dimensional space related to the linguistic context. The Mel-spectrogram was used as input and the output vector was a concatenation of one-hot encoded phonetic identities defined in the ARPABET phoneset (used in the CMU pronunciation dictionary~\cite{cmu2000cmudict}) which is appropriate for the American English speech data. The resulting vector $[\boldsymbol{q_{c1}},\boldsymbol{q_{c2}},\boldsymbol{q_v}]$ representing a V, CV or CCV syllable onset is 64-dimensional -- two sub-vectors encoding 24 consonants (including absence) and one representing 16 vowels. Evaluating this encoder on the Librispeech test set by converting the output to a categorical form results in a recognition rate of 73\%.

For speech synthesis, \emph{VTL}\footnote{Version 2.3 available at \url{https://www.vocaltractlab.de}} was used to realise articulatory targets with the ``JD2'' male speaker and geometric glottis model~\cite{birkholz2019geomglot}. Since we were focused on investigating the upper vocal tract parameters, the glottal parameters were kept constant at the appropriate preset values for the particular segment (e.g. ``modal voice'' for the vowel), with the exception of the \emph{chink area} and \emph{relative amplitude} which were free to be optimised to allow control of the voice onset time. All of the upper vocal tract parameters (Table~\ref{tab:articdims}) were free for optimisation, except the \emph{velum opening} (VO) which was kept closed and the \emph{tongue root} (TRX, TRY) parameters which were derived from the tongue body values~\cite{krug2022efficient}. Timing in the target-approximation trajectories was controlled by two free parameters, one time constant each for the glottal and upper vocal tract parameters.

The somatosensory objectives relied on the same \emph{VTL} configuration to provide proprioceptive or tactile feedback by means of the \emph{tube areas function}. Two simple objectives were defined and represented with values in the range $[0, 1]$. (1) The \emph{vocal tract closure} objective value is 0 when a vocal tract closure is required, e.g. for plosive consonants, and 1 when a minimum opening is required, e.g. for vowels. (2) The \emph{lip closure} objective value is 0 when a closure is formed by the lips and 1 when the lips are open. This is motivated by visual information and was only applied to the first consonant ($C_1$) depending on whether it is a bilabial or other type of plosive. No somatosensory objectives were applied to the intermediate consonant ($C_2$).

A single regularisation objective was implemented by quantifying the coarticulation between any two articulatory targets as the normalised $L_1$ distance between the range-normalised upper vocal tract vectors $\boldsymbol{\tilde{u}}$:
\begin{equation}
  \label{eq:artdist}
  N^{-1}||\boldsymbol{\tilde{u}_1} - \boldsymbol{\tilde{u}_2}||_1, \text{ where }\tilde{u}_{1_i} \text{ and } \tilde{u}_{2_i} \in [0, 1]
\end{equation}
with $N$ the number of upper vocal tract dimensions. In Section~\ref{sec:res} we specifically compare systems with and without this \emph{coarticulation objective} between each consonant and the vowel. Finally, all the relevant articulatory objectives were concatenated into a single vector $\boldsymbol{q_a}$ and combined with the auditory percept to form the optimisation goal $[\boldsymbol{q_a},\boldsymbol{q_{c1}},\boldsymbol{q_{c2}},\boldsymbol{q_v}]$.

To implement the optimisation algorithm~\cite{bergstra2011hpopt} we used the \emph{hyperopt}\footnote{\url{https://github.com/hyperopt/hyperopt} (v0.2.5)} software package. The articulatory space was defined by the speaker model and initially sampled uniformly. The loss function was defined as the weighted sum of the Euclidean distances calculated for the individual sub-vectors, with auditory components having a weight ratio of 2:1 to the articulatory component. For each distinct syllable an ideal objective vector was constructed based on its phonetic constituents. The optimisation algorithm samples articulatory targets which are synthesised by \emph{VTL} and each sample is evaluated by the auditory perceptual mapping and \emph{VTL} tube areas function to determine the resulting vector and associated loss. To improve the computational efficiency of the process, the synthesis of speech and auditory evaluation is only performed when the somatosensory objectives are satisfied. In the case of failure to achieve these objectives, the loss function is set to an arbitrary large value proportional to the loss associated with $\boldsymbol{q_a}$.

\begin{figure*}[t]
    \centering
    \includegraphics[clip,width=1.0\columnwidth]{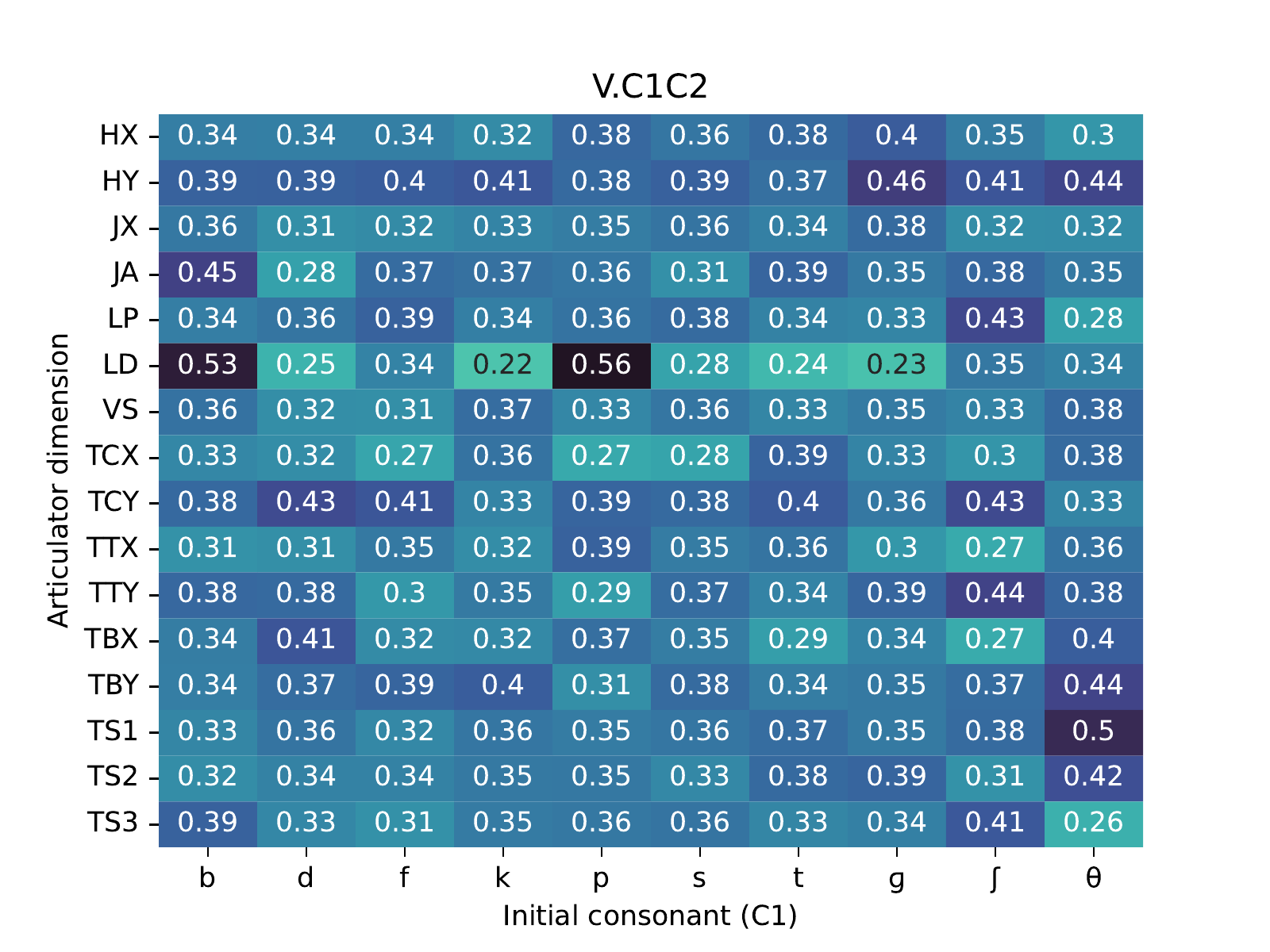}
    \includegraphics[clip,width=1.0\columnwidth]{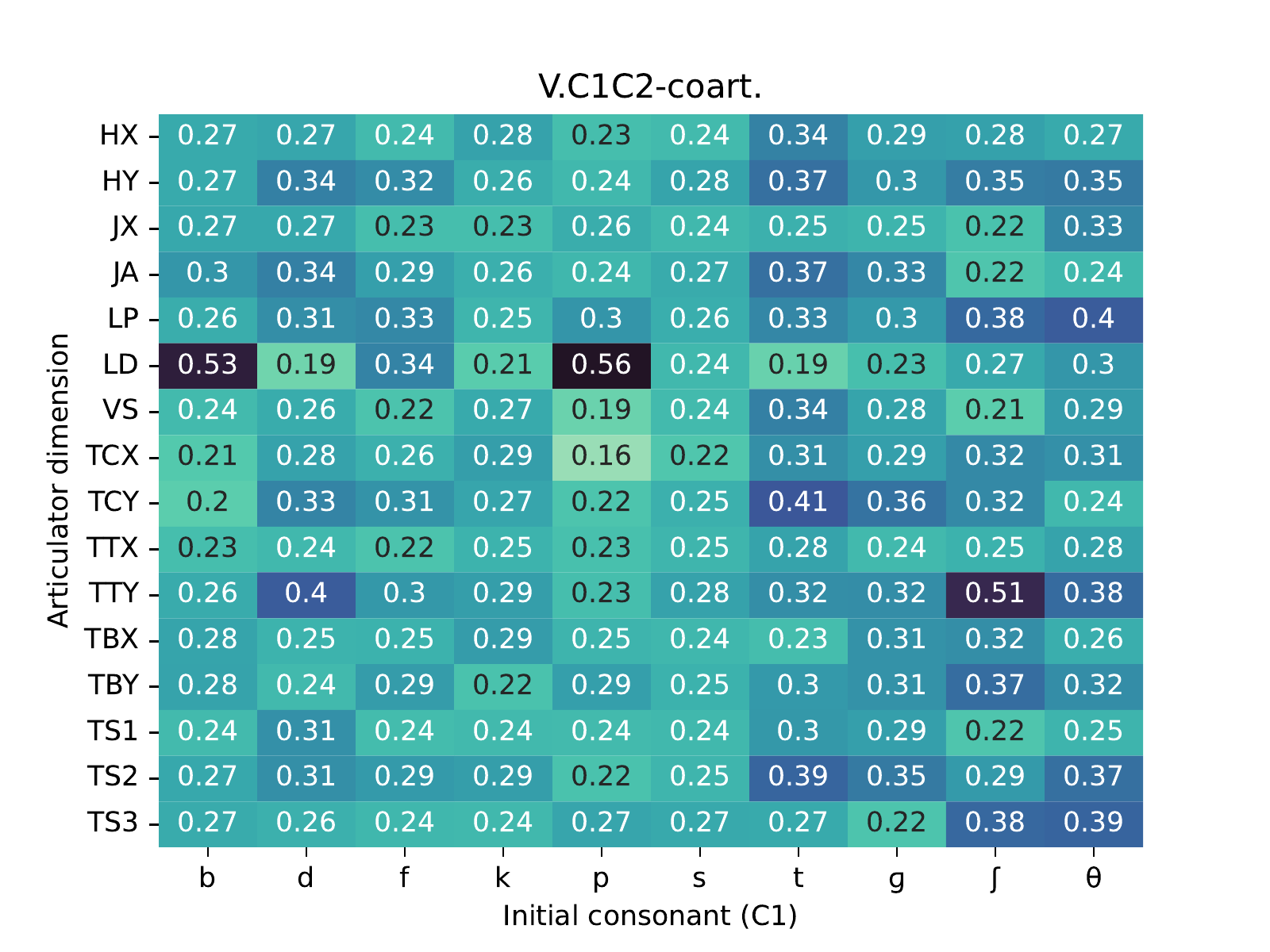}
    \caption{Mean articulatory distances (Eq.~\ref{eq:artdist}) between $C_1$ and $V$ for the $V.C_1C_2$ condition with/without the coarticulation objective.}
    \label{fig:coart}
\end{figure*}

\subsection{Evaluation}
\label{sec:exp:eval}

The proposed framework was evaluated through an experiment designed to investigate the following aspects:
\begin{enumerate}
    \item \emph{Exploration strategies} for complex syllable onsets: Is it necessary to optimise certain segment targets jointly or can this be done independently and in what sequence?
    \item \emph{Coarticulation}: Can we make use of the regularisation objective defined above to reproduce natural observations associated with intra-syllable coarticulation~\cite{liu2021sylsync}?
    \item \emph{Sufficiency}: What is the relative success rate of the process for the range of CCV syllable types occurring in American English and what are the implications for future work?
\end{enumerate}

Since the process is non-deterministic, dependent on random initial exploration, we estimated the success rate using independent repeated trials. To allow for comparison of different syllable types (aspect 3) we set up 5 trials for each of the 150 valid combinations\footnote{Determined by the existence of entries in the CMU dictionary and the LibriSpeech corpus.} of the following sets of segments: $C_1 \in$~\textipa{/b,d,f,g,k,p,s,S,t,T/}, $C_2 \in$~\textipa{/k,l,p,\*r,t,w/}, and $V \in$~\textipa{/A,\ae,2,E,O,I,i:,U,u:/}; a total of 750 independent trials for each experimental setup. Each trial was allocated 5000 iterations leading to a synthesised utterance, i.e. excluding articulatory targets that do not satisfy the basic somatosensory objectives. Four different exploration strategies were investigated (aspect 1):
\begin{enumerate}
    \item \emph{Single-pass, joint} ($C_1C_2V$): Find all segment targets jointly and select the best sample after 5000 evaluations.
    \item \emph{Two-pass, vowel then onset} ($V.C_1C_2$): Find the vowel targets by producing vowel-only utterances in a first pass, then find the onset consonants jointly by producing CCV utterances using the best vowel targets from the first pass.\footnote{To allow for varying coordination requirements in different contexts, the glottal parameters and time constants are never fixed but re-optimised during each pass.} The number of iterations are allocated to the two passes in the ratio 1:4.
    \item \emph{Three-pass, with $C_1$ first} ($V.C_1.C_2$): Find the best targets for each segment by producing $V$, $C_1V$ and $C_1C_2V$ utterances in respective passes (iterations are allocated in the ratio 1:2:2).
    \item \emph{Three-pass, with $C_2$ first} ($V.C_2.C_1$): As in the previous configuration, but with the intermediate consonant explored first.
\end{enumerate}

Each of the above strategies were implemented with and without the coarticulation objective (aspect 2) and the best outcome from each trial was evaluated by the syllable encoder. This was done by mapping perceptual representations to symbols using the \emph{argmax} operation on each sub-vector $[\boldsymbol{q_{c1}},\boldsymbol{q_{c2}},\boldsymbol{q_v}]$ and calculating the identification rate.

\section{Results}
\label{sec:res}

The overall results are summarised in Table~\ref{tab:overall} in terms of identification rates, from which we note the following: (1) There is no significant difference\footnote{We used Welch's unequal variances t-test with a significance level of 5\% throughout.} in the auditory success rate when comparing conditions with and without the coarticulation objective. This means that adding the additional articulatory objective has no negative effect on the goal of optimising for auditory perception. (2) In both cases exploring $C_1$ last on the basis of a pre-optimised $C_2V$ utterance leads to significantly worse results (underlined). (3) When applying the coarticulation objective, jointly optimising the vowel and consonants results in significantly worse outcomes for the vowel ($C_1C_2V$ underlined) and optimising the consonants jointly after the vowel results in a significantly better outcome for syllables ($V.C_1C_2$ in bold).

By repeating this analysis over onsets and vowels we find that the outcomes with and without coarticulation in terms of auditory identification rate are similar in all contexts. In general, all contexts, except \textipa{/d\*r, dw, t\*r, tw/} for onsets and \textipa{/U,i:/} for vowels, have identification rates in excess of 80\%, indicating that those cases are particularly difficult to discover.

To confirm that the coarticulation objective has the intended effect, articulatory distances are visualised in Figure~\ref{fig:coart}. We see that: (1) There are significant reductions in the distance between $C_1$ and $V$ of some articulatory parameters in each case. (2) Some expected patterns of articulatory overlap emerge, e.g. the bilabial targets have lower distances to the vowel in many dimensions except for the \emph{lip distance} (LD).

\section{Discussion}
\label{sec:disc}

Since the optimisation process is partially dependent on the auditory perceptual mapping we are interested in (Table~\ref{tab:overall}), the results should be interpreted carefully. The absolute identification rates are not directly comparable with the recognition rates obtained on natural speech. Instead, we make use of the results in the following ways: (1) To compare the relative success of different exploration strategies in terms of the auditory objectives. (2) To indicate problematic contexts that require further work. The difficult cases listed in Section~\ref{sec:res} could be addressed in future by additional somatosensory objectives or more ecologically plausible articulatory sampling in the case of the onsets and better modelling of duration in the case of the vowels.

The current optimisation-based simulation of babbling is a first step in learning articulation without manual intervention and will be supplemented by a gradient-based learning process towards fluent articulation in future. We invite the interested reader to listen to the samples available at \url{https://github.com/danielshaps/evoclearn_optccv_2022} to get a qualitative sense for the extent to which these correspond to actual babbling utterances.

\section{Conclusion}
\label{sec:disc}

We have presented a flexible simulation of babbling that consists of auditory perceptual, somatosensory and regularisation objectives and demonstrated that it can discover the articulation of syllables with complex onsets. This framework was used to compare different exploration strategies leading to an effective process where the vowel targets are found independently in a first pass and the consonants in the onset are jointly optimised using the vowel as an ``anchor''. With this two-pass procedure it is possible to apply the coarticulation objective (Eq.~\ref{eq:artdist}) without negatively affecting the outcomes in terms of the auditory perceptual goals. This means that the framework and analysis presented in Figure~\ref{fig:coart} can be used to discover the relative (in)dependence of articulatory dimensions in different contexts automatically -- parameter tying was done manually in previous work~\cite{xu2019icphs, vniekerk2020cvopt}. Furthermore, this form of regularisation could prove useful if the utterances generated here are used as a basis for learning forward and inverse models of articulation~\cite{jordan1992distal}.

\section{Acknowledgements}

This work has been funded by the Leverhulme Trust Research Project Grant RPG-2019-241: ``High quality simulation of early vocal learning''.

\bibliographystyle{IEEEtran}

\bibliography{mybib}

\end{document}